# The Health State Function, the Force of Mortality and other Characteristics resulting from the First Exit Time Theory applied to Life Table Data

## Christos H Skiadas

Technical University of Crete
Data analysis and forecasting laboratory
Chania, Crete, Greece.
E-mail: skiadas@cmsim.net

Abstract: In this paper we summarize the main parts of the first exit time theory developed in connection to the life table data and the resulting theoretical and applied issues. Several new tools arise from the development of this theory and especially the Health State Function and some important characteristics of this function. Special attention has being done in the presentation of the health state function along with the well established theory for the Force of Mortality and the related applications as are the life tables and the estimation of life expectancies.
A main part of this work is the formulation of the appropriate non-linear analysis program including a model which provides an almost perfect fit to life table data. This model, proposed in 1995 is now expanded as to include the mortality excess for the age group from 15-30 years. A version of the program is given in Excel and provided at the website: http://www.cmsim.net

## 1. Introduction

### 1.1. Extracting Information from Population-Mortality Data

Over the centuries the systematic gathering of information for births-deaths from the Countries and the related Bureau of the Census or Statistical Agencies gave rise to theoretical and applied studies both qualitative and quantitative in the areas of demography, probability and statistics, applied mathematics and more recently in computer studies and simulations.

The main task was in extracting as much information as possible from the existing data sets but also developing and proposing a framework for the more effective gathering, keeping and providing population and mortality data. To this end today we have quite reliable data sets by based on the theoretical background developed last centuries. Providing reliable data, systematically developed, is the basis of good qualitative and quantitative studies. For birth-death data related to a Country the data sets are quite large assuring that these data include in a statistical reliable way the main part of the information for the status of the population of the specific country over time and age group. These types of data sets include and describe quantitatively the course of the health state of the population including the births and deaths. To a first view the data

---

are simply sets for births and deaths. However, as these data systematically provide complete information for the status of the population by age group or by year of age, it is not so difficult to develop mathematical-statistical techniques to provide information for the health state or for the mortality of the population. From the beginning of the theoretical approaches back to Graunt (1662), Halley (1693) and others four centuries ago, the task was to develop a theoretical framework under the title "Life Tables" facing mortality from the optimistic point of view. The level for the health state of the population emerged indirectly by estimating life expectancies and the life expectancy at birth. Halley commended on a term related to the level of the health state called "vitality" to describe the gradual deterioration of the human health in the course of time. However, the main term accepted, used and applied was the "force of mortality" per age. It was not evident on how it was possible to propose and estimate a similar term for the state of the human health or for the vitality of the organism. It is surprising because the health state is the *direct dominating effect* during the life time and not the inverse called force of mortality.

The health state or the vitality of an organism can be estimated from the birth-death data sets. Bur how? Qualitatively it is a common sense that a good health state of a population will result in higher life expectancy. Quantitatively the answer should be included into the birth-death data and mainly into the death data sets per age or per year of age.

But how to quantify and model the health state of a population by providing a health state function per age? The cause of the lack of developing a quantitative theory is mainly due to the fact that the health state of an individual is a stochastic process and the death is the end of this process when the health state is below a limit called a barrier in the first exit time or hitting time stochastic theory. The theory of stochastic processes was developed only the last century and a well established theory for the first exit time processes was proposed only during last decades. Instead the force of mortality was based on simple probabilistic arguments developed last four centuries.

Another point is that a well established and developed theory for the force of mortality originated from Gompertz and the related theory based on his paper is widely spread and used in demography and actuarial studies and applications. Why to use a new term and a new theory based on the health state of a population and the related function? To be viable and to have a wide spread and applicability a new theory on the health state of a population needs to fulfill the following:

1. To be based on a sound theoretical background.
2. To include the previous theoretical achievements and especially the functions proposed for the force of mortality and the models proposed.
3. To include functions and methods improving the life expectancy estimates.
4. To provide new tools and especially tools related to the estimation of the health state of a population. These tools should include the estimates for the "age of the maximum health state" and the "level" of this health state.

5. To be given with the needed software as to be able the immediate application and testing.

For several years we have worked on exploring the health state process and to provide the material in order to give a quite strong scientific tool useful to theoreticians and practitioners.

In this paper we present the main parts of the theory followed by direct applications to data sets.

## 1.2. Theoretical background

It is expected that a probability density function expressing the death process could arise by using the stochastic theory related to the first exit time, by means of estimating the probability density function concerning the first time for a stochastic process to cross a barrier. The related theory originates from the pioneering work of E. Schrödinger and M. Smoluchowsky from 1915. For the case of the death density function the theoretical attempts where based on "vitality" a term originated from Halley (1693), Gompertz (1825) and few decades ago B. L. Strehler and A.S. Mildvan (1960). Death arises as a cause of the loss of vitality or health which can be considered as a stochastic process. Such a process can be modeled by a simple stochastic differential equation of the form

$$dS_t = \mu_t^* dt + \sigma_t dW_t \qquad (1)$$

Where $S_t$ is the health state or the vitality of the individual, $\mu_t^*$ is a function of the age $t$ and $\sigma_t$ is the diffusion coefficient. In order to avoid confusion with the force of mortality denoted by $\mu$ in the actuarial science we have set the $\mu^*$ for the different function expressing the loss of vitality or the rate of decrease of the health state.

If we assume that $\mu_t^*$ and $\sigma_t$ are independent from $S_t$ the solution of (1) is immediate by integration.

$$S_t = \int_0^t \mu_s^* ds + \int_0^t \sigma_s dW_s \ . \qquad (2)$$

We set

$$\mu_t^* = dH_t / dt , \qquad (3)$$

where $H_t$ is the health state function.

The main problem here is not to find the solution of (1) but the transition probability density function. From (1) we can pass to the associated Fokker-Planck partial differential equation:

$$\frac{\partial p(S_t,t)}{\partial t} = -\mu_t^* \frac{\partial p(S_t,t)}{\partial S_t} + \frac{\sigma_t^2}{2} \frac{\partial^2 p(S_t,t)}{\partial S_t^2} \qquad (4)$$

The solution is given if Janssen and Skiadas (1995) and is of the form

$$p(t) = \frac{1}{\left[2\pi \int_0^t \sigma_s^2 ds\right]^{1/2}} e^{-\frac{(H_t)^2}{2\int_0^t \sigma_s^2 ds}} \tag{5}$$

Whereas for constant $\sigma$ takes the form

$$p(t) = \frac{1}{\sigma\sqrt{2\pi t}} e^{-\frac{(H_t)^2}{2\sigma^2 t}} \tag{6}$$

## 2. The First Exit Time Density Function

The finding of a density function expressing the distribution of the first exit time of particles escaping from a boundary is due to Schrödinger (1915) and Smoluchowsky (1915) in two papers published independently in the same journal issue. Later on Siegert (1951) gave an interpretation closer to our modern notation whereas Jennen (1985), Lerche (1986) and Jennen and Lerche (1981) gave the most interesting first exit density function form. For the simple case presented earlier (6) the proposed form is:

$$g(t) = \frac{|a|}{t} p(a,t) = \frac{|a|}{\sigma\sqrt{2\pi t^3}} e^{-\frac{a^2}{2\sigma^2 t}} \tag{7}$$

Jennen (1985) proposed a more general form for the case of curved boundary using a tangent approximation of the first exit density. Application of this theory to the mortality modeling leads to the following form (earlier work can be found in Janssen and Skiadas (1995) and Skiadas and Skiadas (2007)):

$$g(t) = \frac{|H_t - tH_t'|}{t} p(t) = \frac{|H_t - tH_t'|}{\sigma\sqrt{2\pi t^3}} e^{-\frac{(H_t)^2}{2\sigma^2 t}} \tag{8}$$

As we can see from the last form the term $\left|H_t - tH'_t\right|$ accounts for a local linearization and can be considered as constant in several cases. In this case the simpler form arises:

$$g(t) = \frac{k}{t} p(t) = \frac{k}{\sigma\sqrt{2\pi t^3}} e^{-\frac{(H_t)^2}{2\sigma^2 t}} \qquad (9)$$

That it remains is to find the form of the health state function $H_t$ expressing the loss of vitality process by means that it must be a declining function at the end of the life course. The simpler form is expressed by (Skiadas, 2010)

$$H_t = l - (bt)^c \qquad (10)$$

Where $l$ and $c$ are parameters. When $c=1$ the resulting probability density function is the so-called Inverse Gaussian

$$g(t) = \frac{k}{\sigma\sqrt{2\pi t^3}} e^{-\frac{(l-bt)^2}{2\sigma^2 t}} \qquad (11)$$

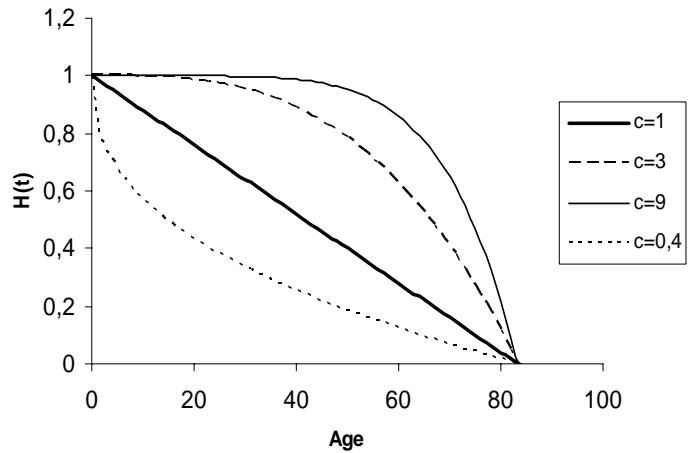

Fig. 1. Simple health state models

The effect of the health state decline modeling is illustrated in Figure 1. The linear approach ($c=1$) is possible only for a very simple system or organism. The death process is expressed by a gradual deterioration of the system without repairing mechanisms. As the organism or the system becomes more complicated and repair mechanisms are present the parameter $c$ is getting larger thus modeling an organism which remains in a good condition for a large period of the life span. These cases are expressed by $c=3$ and $c=9$. The case with $c=0.4$ accounts for a system not well operating.

**Systems Modeling**

Whereas the previous case has the advantage to include the Inverse Gaussian as a special case another approach is also quite important. The main idea is to propose a function modeling the tendency of a living organism to improve its health state during the first period of the life span. This is possible in the case of many complex living organisms by observing that the health state starts from very low values at birth and then is improved considerably until a level and then declines as it happens in human beings. This idea was the basis of a five parameter model proposed by Janssen and Skiadas (1995).

The method for finding the model followed by a system, during a growth or decline process, is described in Skiadas (2008). The basic idea is to construct the $(G(t), g(t))$ diagram, where in the case studied here $g(t)$ is the probability density function and $G(t)$ is the probability function or the cumulative function of $g(t)$.

The data for the probability density function $g(t)$ are given from the statistical bureaus or from the human mortality databases. We normalize these data as to be $\sum_{0}^{n} g(t) = 1$ and we estimate the probability function from $G(t) = \sum_{0}^{n} (g(t)$

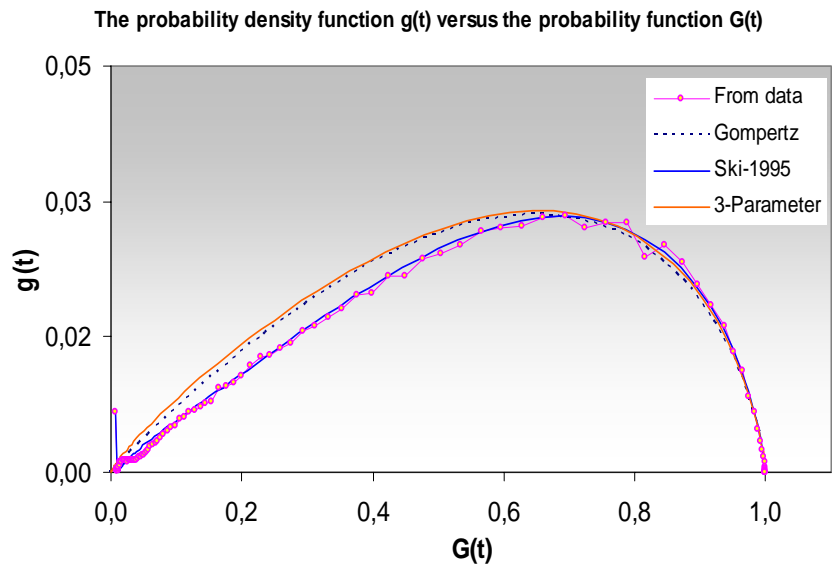

Fig. 2. Modeling the life table system (USA males, 2007)

A first attempt to find a function expressing the growth of a living organism expressed by a magnitude $G(t)$ and a first derivative $g(t)$ can be based on a Logistic growth of the form $g(t) = bG(t)(1-G(t))$, where $b$ is the growth coefficient. The logistic growth is usually not realistic assuming a symmetric form for the growth process. Furthermore there is not a controlling mechanism related to the internal forces of the system. For a birth and death process the controlling mechanism should be related to the relative rate of change of the system and the simpler cases are of the form:

$$Contr1 = \sum_0^n \frac{(G(t+1)-G(t))}{G(t)} \xrightarrow{\lim} -\int \frac{dG(t)}{G(t)} = -\ln(G(t))$$

$$Contr2 = \sum_0^n \frac{(G(t+1)-G(t))}{1-G(t)} \xrightarrow{\lim} -\int \frac{d(1-G(t))}{1-G(t)} = -\ln(1-G(t))$$

Assuming a system governed by the simple exponential growth $g(t) = bG(t) \cdot Control1$ for the first case and $g(t) = b(1-G(t)) \cdot Control2$ for the second case the Gompertz models follow as the best strategies selected from the internal mechanisms of the system. The related equations have the form:

$g(t) = -bG(t) \cdot \ln(G(t))$, the maximum is at $1/e \approx 0.368$,

and

$g(t) = -b(1-G(t)) \cdot \ln(1-G(t))$, the maximum is at $1-1/e \approx 0.632$

These are the two forms of the so-called Gompertz model. The first is used in statistics and the second is popular in demography and in actuarial science. The second case is precisely the form which fits to life table data sets. This is presented in Figure 2. The data are from the Human Mortality Database for males in USA (2007). From the data set we estimate a maximum $g(t)=0.0326$ for $G(t)=0.693$. The later is relatively close to 0.632 for the Gompertz model. A 3-Parameter model proposed last year (Skiadas, Jan 2011) provides a maximum $g(t)=0.0335$ for $G(t)=0.643$ and is slightly better than the Gompertz. However the behavior of the Gompertz and the 3-Parameter model during the first and middle stages of the growth process diverges from the curve produced from the data sets. The best model is the Ski-1995 model with maximum $g(t)=0.0328$ for $G(t)=0.691$, values almost identical to the estimated from the data.

The 3-Parameter model has the form: $g(t) = k(c-1)(bt)^c (t)^{-3/2} e^{-\frac{(bt)^{2c}}{2t}}$,

and the Ski-1995 has the form:

$g(t) = \frac{k}{\sqrt{t^3}} e^{-\frac{(H_t)^2}{2t}}$, where the health state function $H_t$ is

$$H_t = a_1 + at^4 - b\sqrt{t} + lt^2 - ct^3$$

This model is developed in order to include the infant mortality ($a_1$, $a$, $b$, $l$ and $c$ are parameters).

### 3. A special property of the Health State Function *H*(*t*)

As we have introduced the health state as the opposite of mortality and the health state function as the opposite of mortality function we should find a simple by quite effective method to estimate *H*(*t*) from Life Tables or from Mortality Data. From the first introduction of life tables it was evident that the optimistic term "life" was preferred than the pessimistic "mortality". Gompertz (1825) commended "life contingencies" in his seminal paper "On the Nature of the Function Expressive of the Law of Human Mortality, and on a New Mode of Determining the Value of Life Contingencies". Halley (1693) estimated mortality in his paper on "An Estimate of the Degrees of the Mortality of Mankind, drawn from curious Tables of the Births and Funerals at the City of Breslaw; with an Attempt to ascertain the Price of Annuities upon Lives". The terms Life Table and Life Table Data are more frequently used especially from actuaries as such terms emphasize the optimistic view of the life course.

As the death data from a country include, as an average, a large part of the information related to the health state of the population it is surprising that no-quantitative works are present related to the health state of the population during the last centuries. We understand that the health state data are included into the mortality data and the main task is to find a method to attract this information from the mortality data or from the life table data. It is clear that both population data and mortality data should contribute as is already the case when constructing life tables.

The theory of the proposed health state function is relatively new. Special futures are expected to arise by continuing study and research. More work is needed especially on finding the form of the unknown function *H*(*t*) and the properties of this function. Turning buck to the general equation proposed (9)

$$g(t) = \frac{k}{t} p(t) = \frac{k}{\sigma\sqrt{2\pi t^3}} e^{-\frac{(H_t)^2}{2\sigma^2 t}},$$

we see that by estimating the parameters of a model with any health state function *H*(*t*) the diffusion coefficient *σ* cannot be estimated explicitly and it will by an internal part of these parameters. However, the stochastic differential equation (1) provides a simple method to estimate *σ* from the data sets. We can immediately observe that

$$(dS_t)^2 = (\sigma_t dW_t)^2 = \sigma_t^2 dt , \qquad (12)$$

and the estimation of a constant *σ* arises from

$$\sigma = \sqrt{\frac{\sum_0^n (S_{t+1} - S_t)^2}{n}} \qquad (13)$$

Without loss of generality we can set $\sigma=1$ in equation (9) for $g(t)$ and continue with the simple form of the model

$$g(t) = \frac{k}{\sqrt{t^3}} e^{-\frac{(H_t)^2}{2t}} \qquad (14)$$

Where the new parameter $k = k^* / \sqrt{2\pi}$

We turn now to have an estimate of the form of the unknown health state function $H(t)$ by rearranging the previous equation and expressing $H(t)$ as a function of $g(t)$. The resulting function is of the form

$$H_t = \left| \left( -2t \ln \frac{g(t)\sqrt{t^3}}{k} \right)^{1/2} \right| \qquad (15)$$

To assure a positive sign in the between brackets term in the right hand side of the last formula the following relation holds

$$k \geq g(t)\sqrt{t^3} \; . \qquad (16)$$

Thus we can immediately have an illustration of the form of the function $H(t)$ by introducing the values for the deaths $g(t)$ per age $t$ from any annual data sets from the human mortality databases in the above formula after estimating the parameter $k$.

Application to the mortality data of United States for 2000 provided by the Human Mortality Database gave $k=31.10$ at 88 years of age for females (Figure 3) and $k=25.49$ at 83 years of age for males (Figure 4).

The very interesting finding is that along with the characteristic value of $k(T)$ we estimate the age of the maximum death rate $T$ from the relation:

$$k(t) = g(t)\sqrt{t^3} \qquad (17)$$

The estimation of parameter $k$ is an easy task if we observe that the function $H(t)$ should be continuous as is the function $g(t)$, it should decrease continuously in the older ages and it should be zero (or very close to zero) at the age of the maximum death rate. As it is illustrated in Figure 3 there is only one curve fulfilling the previous arguments when $k=31.10$ (USA 2000, females) whereas the cases with $k<31.10$ violate continuity (see the curves for $k=5$) and the cases with $k>31.10$ show a gradual increase in the old ages (see the curves for $k=60$). In the same Figure we have indicated the form which should have the unknown function *H*(*t*) (curve with small circles). Similar results are illustrated in the next Figure 4 (USA 2000, males).

The Health State Functions for males and females for the same time period (2000) for USA are illustrated in Figures 5. The *H*(*t*) for females has higher values than for males and the same holds for the age of zero *H*(*t*). In both cases the Maximum Health State is found to be between 30 and 45 years of age, a quite reasonable finding.

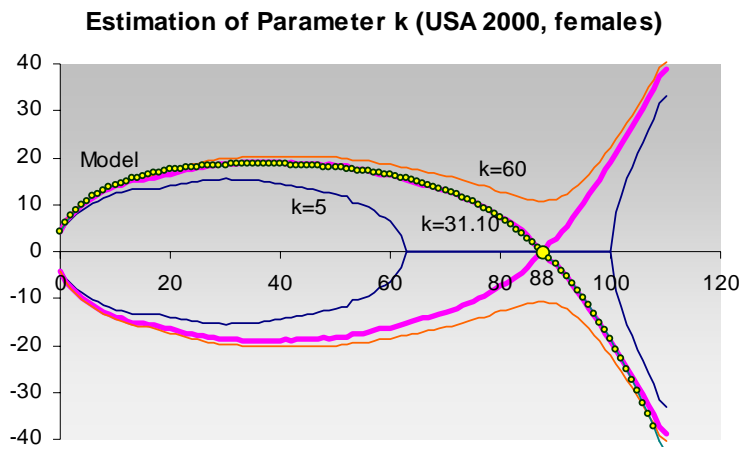

Fig. 3. The parameter *k*(*t*) for females in USA (2000).

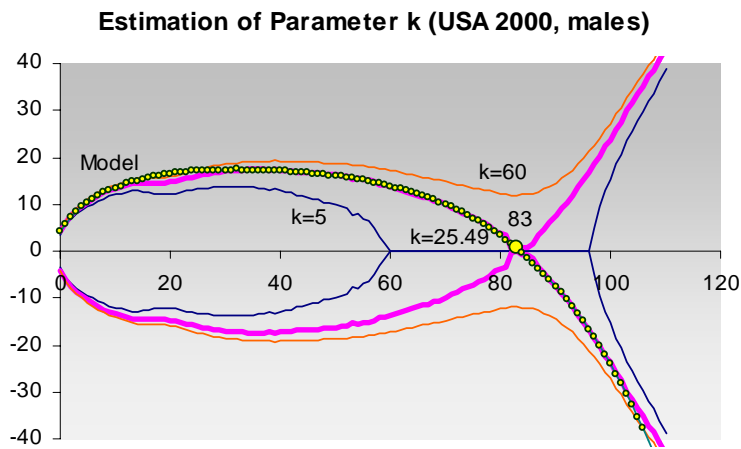

Fig. 4. The parameter *k*(*t*) for males in USA (2000).

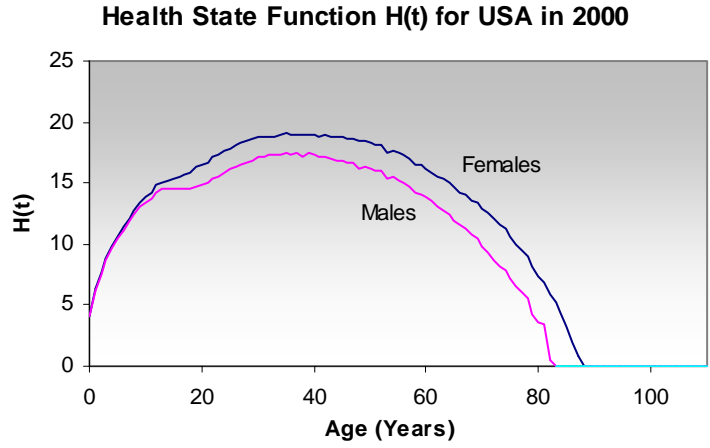

Fig. 5. Health state function for males and females in USA.

## 4. Methodological Issues

A very important point is the selection of the appropriate data sets usually provided by the Bureau of the Census of the specific Country. In this part of the study we mainly use the actuarial notation replacing the time *t* by the years of age denoted by *x* in all the previous notations and equations. The data needed are the number of deaths per age and year ($d_x$) and the observed population size also per age and year ($l_x$). Then the Force of Mortality $\mu_x$ is given by

$$\mu_x = \frac{d_x}{l_x} \qquad (18)$$

When the data are provided in terms of $\mu_x$ we proceed to find the $H_x$ directly. As in our calculations we need data for estimating the probability density function $g(x)$ we can find this function from the definition of $\mu_x$ that is from the formula

$$\mu_x = \frac{g(x)}{1 - \int_0^x g(t)dt} \qquad (19)$$

After rearranging and differentiating we obtain

$$(\ln(g_x))' = (\ln(\mu_x))' - \mu_x \qquad (20)$$

Rearranging and integrating we obtain

$$\ln(g_x) = \ln(c) + \ln(\mu_x) - \int_0^x \mu_t dt \qquad (21)$$

We can set the integration constant $c=1$ to get the following form for the probability density function:

$$g_x = \mu_x e^{-\int_0^x \mu_t dt} \qquad (22)$$

It should be noted that the resulting values for $g_x$ will be normalized so that

$$\int g_x = 1$$

The formula for $g_x$ provides the probability density function $g(x)$ as a function of $\mu_x$. Accordingly we can find the following form for the Health State Function

$$H_x = \left| \left( -2x \ln \frac{\sqrt{x^3} \mu_x e^{-\int_0^x \mu_t dt}}{k} \right)^{1/2} \right| \qquad (23)$$

The parameter $k$ is now given by

$$k = \sqrt{x^3} \mu_x e^{-\int_0^x \mu_t dt} \qquad (24)$$

Or by the simpler form

$$\ln(k) = \frac{3}{2}\ln(x) + \ln(\mu_x) - \int_0^x \mu_t dt \qquad (25)$$

The resulting graphs of an application to Japan (2008, Females) are presented in Figure 6a. The function $H(x)$ crosses the horizontal axis at the age of 89 years where the curve expressing $k(x)$ and has a maximum. This maximum at $k=40.42$ is the required value for the parameter $k$ so that the function $H(x)$ exists, is continuous and declines at the old ages.

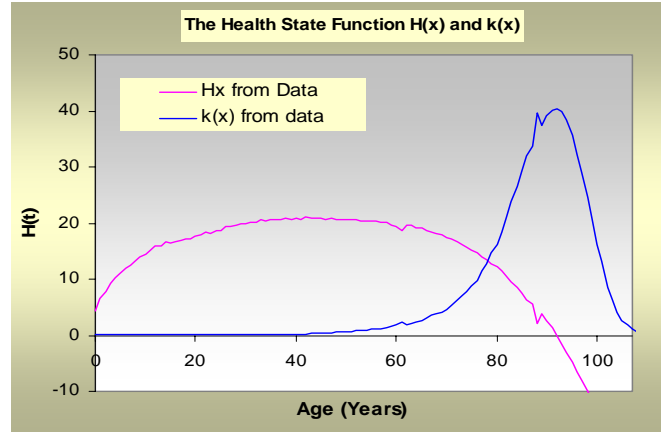

Fig. 6a. Health state function and *k(x)* for Japan (2008, females)

The resulting graphs for *H(x)* provide a non-symmetric parabola like form. This function describes the mean value of the health state of the population for a specific year as a function of age. The health state starts from a low level at birth and then it continues growing until the ages between 30 and 45 and then declines to zero at the age of the maximum death rate.

As we have observed a parabola like form for the *H(x)* function we can find the age *x* where *H(x)* receives its maximum. For the case of USA in 2000 we found *x*=35 years and *H(x)*=19.07 for females and *x*=35 years and *H(x)*=17.47 for males. Females in Japan show higher health state than females in USA for the year 2000 (see Fig. 6b).

We have thus proposed simple and quite important estimators of the Health State of a population or of the vitality in terms of Halley and others. The new approach emerged from the first exit time theory developed and applied on the mortality data for various countries.

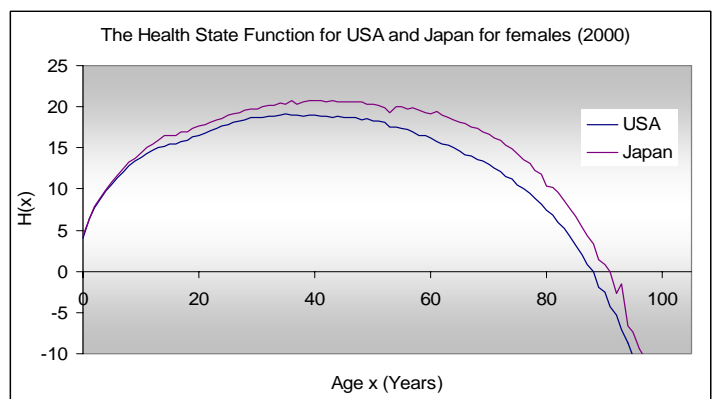

Fig. 6b. The Health State Function in USA and Japan

## 5. The Force of Mortality

The introduction of the force of mortality and the commonly used logarithmic form $\ln(\mu_x)$ have some shortcomings. The point related to the maximum death rate is not present. It is hidden in the almost linear part of the logarithmic form. The same holds for the age of the maximum health state, a period between 30-45 years of age. We have introduced a parallel movement of the logarithmic form for the force of mortality as to pass from the point of the maximum death rate in the horizontal axis (see Figure 7). To achieve this we first observe that the mortality curves are almost linear at the age-periods close to the age of the maximum death rate. The displacement *y* is estimated from the form:

$$y = a + \beta x_{\max} \qquad (26)$$

Where *a* and *β* are estimated from the data provided or from the best fit results depending if the parallel movement accounts for the $\ln(\mu_x)$ from data or the $\ln(\mu_x)$ from the model fit. For the case of USA (males, 2007) the respective values are (*a*=-10.13 and *β*=0.0915) for data and (*a*=-10.15 and *β*=0.0918) for the best fit curve for $\ln(\mu_x)$. The used program is presented at the end of this paper. The applied health state function (Janssen-Skiadas, 1995) has the form:

$$H_x = a_1 + ax^4 - b\sqrt{x} + lx^2 - cx^3 \qquad (27)$$

The curve *k(x)* is very important for estimating the age of the maximum death rate and also for estimating the $k_{max}$ at this age. This curve *k(x)* is also included into the same graph. The estimated values for males in USA (2007) are $k_{max}$=26.43 from data and $k_{max}$=25.88 from fit at 87 years of age.

After estimating the parameter *k* and finding the form of the unknown function *H(x)* that it remains is the proposal and application of functions which best fit the data. It is clear that a polynomial function as the (27) could be a first selection. However we should pay attention to the presence of a loss of the health state level for people from 15 to 30 years of age. To model this case extra parameters are needed turning the models to be more "heavy" and difficult to apply.

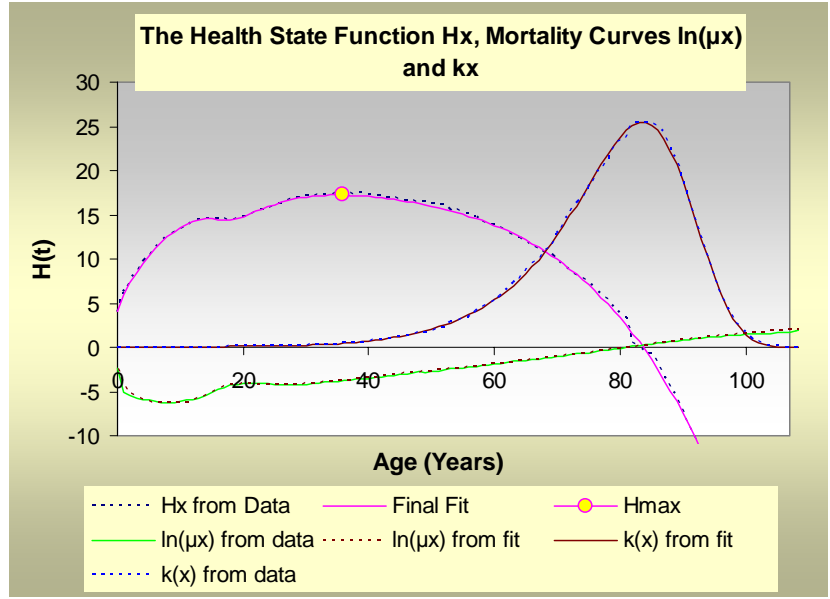

Fig. 7. The health state, the mortality and the *k(x)* functions for USA 2007, males.

One simple modeling technique is first to apply a polynomial like form (27) for *H*(*x*) and, after the first parameter estimate, to add an extra term for the estimation of the non-symmetric health state case for the interval from 15 to 30 years of age. For the later case we introduce and apply a Gompertz model to the data expressing the differences between the original data and the estimated. The selected Gompertz model has the form

$$g_1(x) = k_1 e^{-b_1 x} e^{-l_1 e^{-b_1 x}} \qquad (28)$$

The main advantage of the method proposed is that we have an explicit function providing the distribution of deaths per year of age of the population and then we easily calculate the mortality curve and other characteristics and of course to construct a life table from the fitting curve and calculate the life expectancies. The inverse problem to estimate the mortality curve and then the death distribution faces difficulties. From a point of view it is complicated to transform the death data by taking logarithms, to estimate the mortality curve and then to come back to re-calculate the death distribution. Another important future of this method is that we can easily extract the health state and the force of mortality only from death data per year and age provided that the birth-death system is in equilibrium in Keyfitz (2005) terms.

## 6. The Excel Program

To simplify things we have introduced all the necessary models and estimation techniques in Excel so that the user can make relatively easy the calculations for any specific case.

First we introduce in the P column of the Excel spreadsheet (see Figure 11) the Death data and in the Q column the Population data as provided by the Human Mortality Database. The indicator in A-B 24 cell is set to 1. If only Death data are introduced it is set to 0. Then the program estimates the parameters by using a non-linear regression analysis following a Levenberg(1944)–Marquardt(1963) like algorithmic method (Skiadas, 1986). The resulting parameters are illustrated in B-C 12 to B-C17 cells. As the method requires the introduction of initial parameters the program already includes a first guess for these parameters in the red colored cells (C2, B6, C7, D8, E9 and F10). For the parameter *k* in cell D8 the first guess is already very close to the final estimates as it is calculated from the original data set by based on the formula: $k(x) = g(x)\sqrt{x^3}$ . You also have to introduce the number of the data points in cell F4 and to select in C3 the appropriate step for the algorithmic method used. It should be between 0 and 1. The parameter selection gives a quite good fit as it is illustrated in Figure 8 (green line) but without including the mortality excess for the years 15 to 30. Then we introduce an appropriate Gompertz function to correct the first estimates. This is done semi-automatically by the program; the user should only change *a*1 in cell C2 as to minimize the Sum of Squared errors in cell D3, for minimizing the errors between the mortality curves (from data and estimated). The sum of squared errors between the death distribution curves (from data and estimated) are given in cell D4 as is illustrated in Figure 11. The fitting is almost perfect giving the opportunity for applications in various countries and for large time periods.

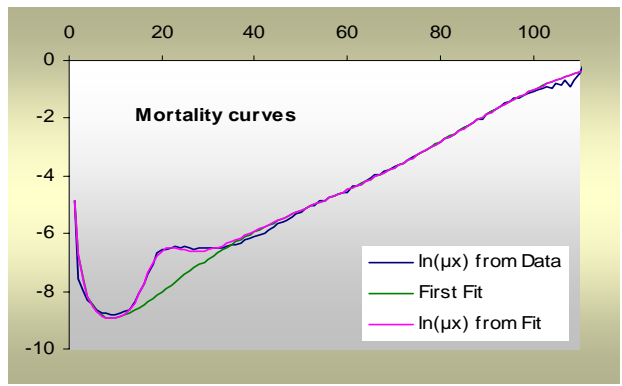

Fig. 8. Mortality curves (First fit, final fit and from data)

Some of the futures from the new method used are:

The program provides a graph of the Health State Function and the related maximum value and of the Mortality Function (from data and estimated, first estimate and final).

An estimate for the curvature of $H(t)$ is also provided along with the maximum deterioration age.

Two Life Tables are provided in order to estimate the Life Expectancy and the Life Expectancy at Birth by using the provided data and the estimated from the program (best fit). The estimates are also provided

A graph providing the probability density function $g(x)$ and the mortality curve $\ln(\mu_x)$ versus the probability function $G(x)$.

In short any information needed from statisticians, actuaries, demographers and any other interested scientist or practitioner is provided including the death distribution (Figure 9) and the generated stochastic paths based on the health state function and the estimated diffusion coefficient $\sigma$ (Figure 10). The estimated value of the diffusion coefficient is $\sigma=1.257$ for USA 2007 (males).

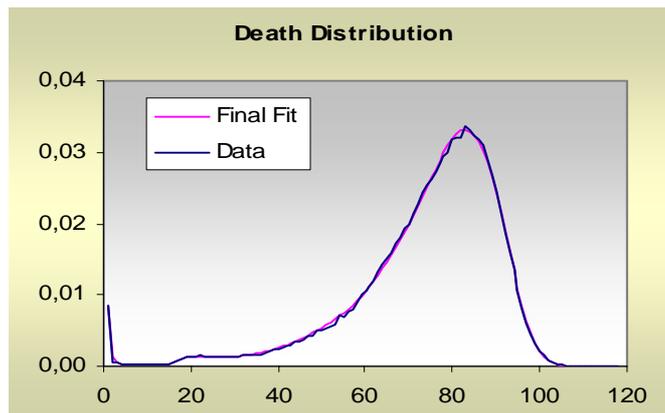

Fig. 9. The death density function.

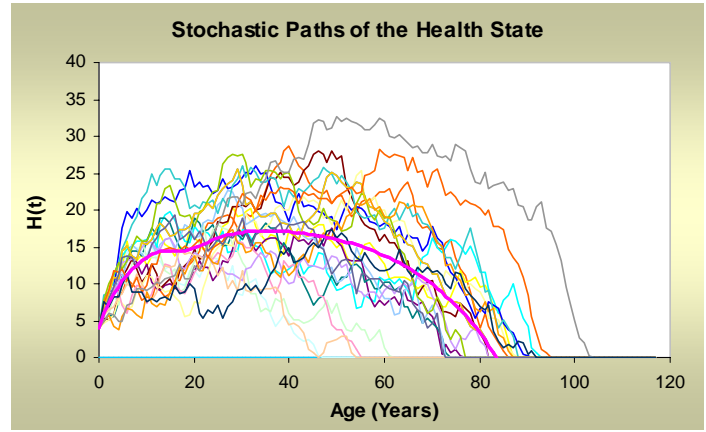

Fig. 10. The health state function and stochastic paths

## 7. Conclusions

We have introduced and analyzed a general theory of health state of a population directly related to mortality and the force of mortality and various estimates included in statistics and actuarial science. The aim was towards of providing tools for measuring the health state of a population while keeping in the same framework the existing tools and especially the force of mortality and the related applications in developing the life tables. By using this theory and the provided tools and the Excel programs the estimates and fitting curves are very easy. Special application techniques are introduced and a model with an almost perfect fit is provided with the Excel program. We are now able to classify the various Countries according to the level of the maximum health state for a specific year something very important for various organizations as the World Health Organization.

**Acknowledgements:** With this study we conclude a fascinating exploration into the systems modeling world. Deterministic, Stochastic and Chaotic models, methods and techniques along with Data Analysis and Simulation Methods gave rise to a 30 years works along with the development and launch of the appropriate computer programs. Many colleagues supported this exploration via the reference system or by positive or negative criticisms. I am happy that due to a good data collection, development and restoration system we can, by using computers and computer programs, to give precise and detailed answers and strongly support the proposed models, methods and techniques. The challenge and motivation from the environment of the Technical University of Crete and the Department of Production Engineering and Management were very stimulating. Many thanks to my doctoral students and Charilaos Skiadas.

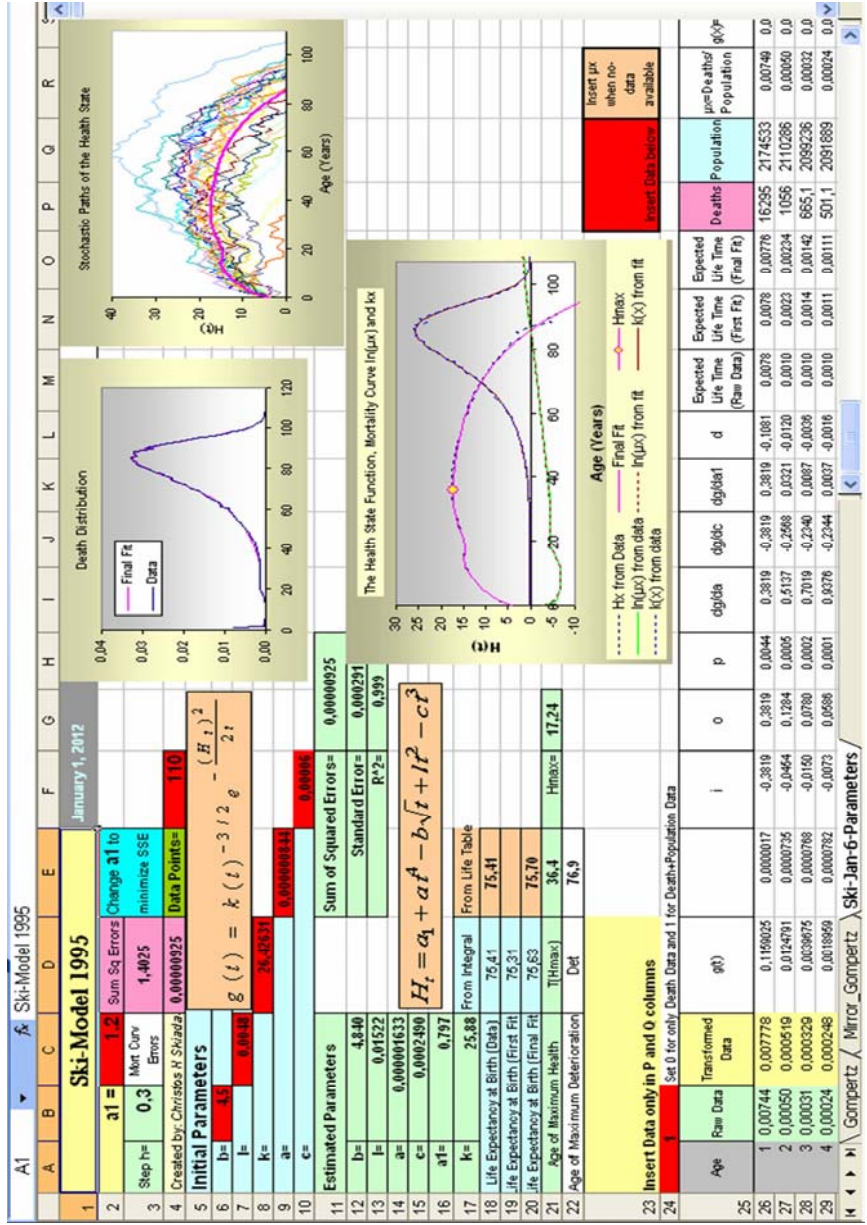

Fig. 11. The main page of the Excel Program.